ECGDeDRDNet: A deep learning-based method for Electrocardiogram noise removal using a double recurrent dense network

Sainan xiao[1,2], Wangdong Yang[1*], Buwen Cao[1,2*], Jintao Wu[1]

1. College of Computer Science and Electronic Engineering, Hunan University, Changsha, Hunan 410082, China
2. College of Information and Electronic Engineering, Hunan City University, Yiyang, Hunan 413002, China

Correspondence: Yangwd@hnu.edu.cn; yongheng@hncu.edu.cn

**Abstract:** Electrocardiogram (ECG) signals are frequently corrupted by noise, such as baseline wander (BW), muscle artifacts (MA), and electrode motion (EM), which significantly degrade their diagnostic utility. To address this issue, we propose ECGDeDRDNet, a deep learning-based **ECG De**noising framework leveraging a **D**ouble **R**ecurrent **D**ense **Net**work architecture. In contrast to traditional approaches, we introduce a double recurrent scheme to enhance information reuse from both ECG waveforms and the estimated clean image. For ECG waveform processing, our basic model employs LSTM layers cascaded with DenseNet blocks. The estimated clean ECG image, obtained by subtracting predicted noise components from the noisy input, is iteratively fed back into the model. This dual recurrent architecture enables comprehensive utilization of both temporal waveform features and spatial image details, leading to more effective noise suppression. Experimental results on the MIT-BIH dataset demonstrate that our method achieves superior performance compared to conventional image denoising methods in terms of PSNR and SSIM while also surpassing classical ECG denoising techniques in both SNR and RMSE.

# 1 Introduction

Electrocardiogram (ECG) signals provide a non-invasive representation of cardiac electrophysiological activity through skin surface recordings [1]. As a critical diagnostic tool, ECG enables:
- detection of cardiovascular disorders [2];
- arrhythmia identification [3,4];

- cardiac function assessment [5];
- treatment monitoring [6,7];
- preoperative evaluation [8,9];
- asymptomatic screening [10,11].

However, ECG signal quality is frequently compromised by various artifacts, including: Electrode motion artifacts (EM) [12, 13]. Baseline wander (BW) [14, 15], and Muscle artifacts (MA) [16, 17]. These artifacts can significantly distort diagnostic features, potentially leading to clinical misinterpretation. Consequently, developing robust denoising methods that preserve clinically relevant characteristics while effectively removing artifacts remains an essential research challenge.

One widely adopted strategy for eliminating noise from electrocardiogram (ECG) data involves decomposing complex signals into several intrinsic mode functions (IMFs) [18-20]. This approach primarily relies on empirical mode decomposition, which presents significant drawbacks; crucial information may be lost in the discarded IMFs, and extracted IMFs might suffer from mode mixing and endpoint effects. An alternative denoising technique is wavelet transform (WT) [21-23], which extends signal representations in both the time and frequency domains using wavelet functions. WT processes involve decomposing or reconstructing signals through wavelet bases formed by scaling and translating these wavelets. WT features excellent localization properties in both time and frequency domains, strong noise reduction capabilities, and broad application scope. However, WT requires manual function selection and exhibits limited adaptability.

In [24, 25], a sparse model-based method was introduced, wherein ECG signals are decomposed into sparse and residual components. The former is utilized for signal evaluation, encompassing tasks such as dictionary generation and signal coefficient decomposition. The principle of sparse representation involves expressing a signal as a linear combination of atoms within a signal sequence, calculating sparse solutions, and subsequently employing these solutions to reconstruct the signal. The sparse model denoising method boasts robust noise removal and generalization capabilities, effectively eliminating various types of noise. Nevertheless, this method faces challenges such as model interpretability, high computational complexity, substantial time costs, and slow convergence rates.

Additionally, another approach [26] focuses on removing BW and

MA noise from ECG signals by converting the signal from the time domain to the frequency domain via Fourier Transform (FT), then filtering out spectral components corresponding to artifacts. After discarding the noise components and performing an inverse FT, a clean signal can be obtained. Although FT facilitates the analysis of temporal and frequency correlations of signals and is extensively used in ECG denoising research, it necessitates analyzing the entire signal, lacking real-time and local analysis capabilities.

Lastly, filter denoising methods [27-29] have been employed to remove noise to some extent. However, their effectiveness is suboptimal, often leading to distortions in the original ECG signal, thereby compromising its medical value and posing a risk of misdiagnosis.

Recently, deep learning (DL) has garnered increasing attention due to advancements in computing resources, continuous innovation in network architectures, and enhanced capabilities for processing large datasets. DL has successfully addressed various technical challenges, including image processing [30, 31], signal processing [32, 33], and specific ECG-related tasks such as arrhythmia classification [34-36], ECG signal segmentation [37, 38], and ECG signal reconstruction [39, 40]. Notable DL models applied to these areas include auto-encoders [41, 42], simple and complex convolutional neural networks (CNNs) with or without residual connections [43, 44], and generative adversarial networks (GANs) [45-47]. These models have also been utilized for noise reduction in ECG signals. Despite significant progress, existing studies exhibit limitations, such as the absence of k-fold cross-validation, limited model generalizability, potential for further performance improvements, and inadequate performance metrics.

Inspired by a method that removes rain streaks from single images using a double recurrent dense network [48], and considering the periodic nature of ECG signals, we propose a DL-based approach for ECG denoising. Our method leverages a double recurrent dense network, drawing insights from image denoising techniques. The architecture of our proposed model is illustrated in Fig. 1. To address optimization challenges, a double recurrent strategy is employed to enhance the reuse of information between noisy and relatively clean ECG segments. Noise removal is performed sequentially, with each stage's output combined with the original noisy ECG signal to serve as input for the subsequent stage. Additionally, a composite loss function incorporating L1 loss, L2 loss, and SSIM loss is designed to ensure effective noise removal. Experimental results demonstrate that our proposed method surpasses traditional denoising techniques for ECG signals.

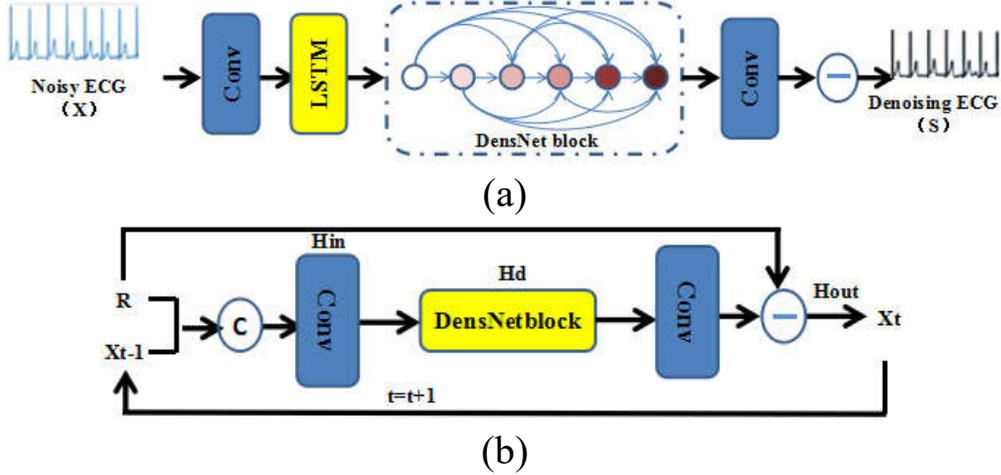

**Fig.1.** The framework of our proposed model

The primary contributions of this paper are summarized as follows:

1) This paper introduces a Double Recurrent Dense Network (ECGDeDRDNet) designed for enhanced information reuse. By integrating Dense Blocks with LSTM units [49], ECGDeDRDNet facilitates the flow of noise-related information, while cascaded stages effectively leverage denoising results from previous stages to improve overall performance.

2) A hybrid loss function combining L1 loss, L2 loss, and SSIM loss is proposed. Experimental results demonstrate that this loss function outperforms existing loss functions in terms of noise reduction efficacy.

3) As a versatile deep network model, our proposed method surpasses traditional wavelet transform (WT) techniques in removing noise from ECG signals.

The remainder of this paper is organized as follows: Section 2 outlines the noise models under consideration. Section 3 details the proposed denoising methodology. Section 4 presents experimental details and provides an analysis of the results, comparing the quantitative and qualitative performance of the proposed model against three different denoising methods. The paper concludes with a summary in Section 5.

## 2 Denoising model

Generally, denoising can be formally defined by the following

equations:
$$X = S + N \quad (1)$$
Where X represents the noisy signals/images, S denotes the clean signals/images, and N is the noise. The objective of the denoising task is to remove as much noise interference N as possible from X, while preserving the useful original signal/image S. This process is mathematically expressed as: as:
$$S = X - N \quad (2)$$
Based on Eq.(2)., many studies have constructed denoising network models under the assumption that noise predominantly resides in the high-frequency components. In these models, ECG images are decomposed into a base layer and a detail layer. The differences between the detail layer with the noise and without noise is learned.

Statistically, most of noise is concentrated in the detail layer, as illustrated in Fig.2. However, Fig.2. also reveals that significant residual noise remains in the base layer, which limits the further improvement in denoising performance.

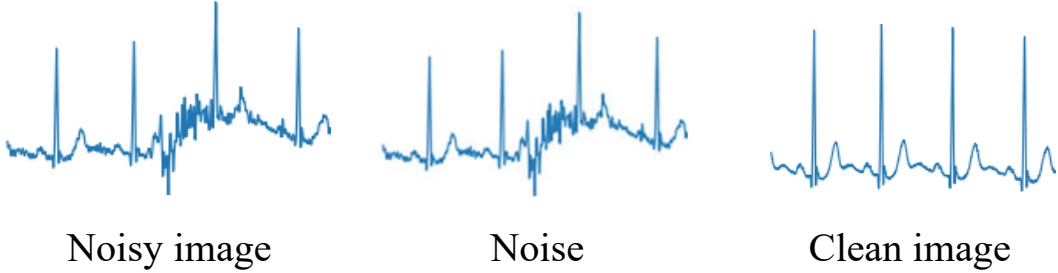

    Noisy image          Noise          Clean image

**Fig.2** A pair of ECG images of the composite data set, ECG image is a combination of clean image and noise.

Instead of focusing on removing noise from the detail layer, our approach restores the clean image directly from the noisy input. Let $D(\omega)$ present the denoising model, where $\omega$ is the input of the model. The goal of denoising is to directly output a clean image from the inputs containing noise.
$$\widehat{N} = D(X) \quad (3)$$
Here, $\widehat{N}$ represents the prediction for the noise $N$, and the model can either directly output the clean image or indirectly output it by estimating the noise. Specifically, $\widehat{B}$ denotes the model's prediction for the clean image $B$.

In practice, we learn the noise prediction from the model and then estimate the clean image by subtracting the predicted noise from the noisy image, as described below:
$$\widehat{B} = X - H(X) \quad (4)$$
Here, $H(X)$ represents the noise estimation function applied to the noisy image $X$. The noise layer can be considered as a combination of noise in multiple directions and other complex noises, such BW, MA, EA, etc. Furthermore, real-world collection can generate additional complexities, such as sudden detachment of the collection device. Therefore, learning the noise pattern is simpler and more effective than directly learning the clean image from the noisy input.

## 3 Denoising method

Smiliar to [48], the architecture of our proposed network involves the following aspects: the Recuuent Dense Network (RDNet), the ECG-specific Double Recurrent Dense Network (ECGDeDRDNet), and the loss function design of ECGDeDRDNet. Based on the concept of the progressive removal of noise, we developed a Recurrent Dense Network model (RDNet). To better adapt this framework for ECG denoising, the Long Short-Term Memory (LSTM) units are integrated with RDNet, resulting in ECGDeDRDNet. This integration leverages the strengths of both recurrent structures and dense connections, thereby significantly enhancing the performance of our denoising model. The synergy between LSTM and RDNet facilitate the use of information from previous putputs to guide subsequent learning stages, improving overall model effectiveness. To achieve optimal denosing performance, a hybrid loss function of $L_1$ loss, $L_2$ loss, and Structural Similarity Index Measure SSIM is designed to ensure the improvements in both quantitative evaluation metrics and visual quality. Specifically, the inclusion of SSIM addresses perceptual quality, while $L_1$ and $L_2$ losses foucs on pixel-wise accuracy and robustness against outliers.

### 3.1 RDNet

The RDNet primarily consists two steps [48]. First, the denoising image (S) is extracted through the network, as illustrated in Fig.1(a). Second, both the noisy image (X) and the denoised image (S) are recurrently fed into the network to progressively remove the noise, as presented in Fig.1(b). As depicted in Fig.1., the RDNet is composed of three main components: the input layer $H_{in}$, the DenseNet [50] layer $H_d$, and the residual output layer $H_{out}$. The input layer $H_{in}$ is used to extract 32 noise features by executing a convolution operation, and the noise features are then passed to DenseNet layer $H_d$, which further promotes feature reuse, enhancing information flow between layers. It is crucial for extracting the noise layer effectively. Fig.2.

illustrates the structure of DenseNet block, which contains 5 layers, each with 32 filter maps followed by a non-linear ReLU function. The output layer $H_{out}$ performs a residual operation by subtracting the noise from the noisy input images. The relative clean image S can be calculated using the following formula:

$$S_t = H_{out}(H_d(H_{in}(S_{t-1}, inputs))) \qquad (5)$$

Here, $inputs$ refer to the noisy images, and $S_{t-1}, S_t$ represent the outputs of the RDNet at the $(t-1)^{th}$ and $t^{th}$ iterations, respectively.

### 3.2 ECGDeDRDNet

LSTMs are explicitly employed to store information. In image processing, iteration is commonly utilized to enhance the performance of network models. When iterating, information from previous iterations is valuable for improving the current iteration's performance [48]. In this study, the LSTM model [51] is introduced to improve the performance of single image denoising. To further enhance the performance of single image denoising, a recurrent structure is integrated into the RDNet to remove the noise in multiple stages, resulting in the ECGDeDRDNet, as presented in Fig.3. This architecture is derived from RDNet and includes additional components located in the input layer $H_{in}$ and the DenseNet layer $H_d$ (not detailed in Fig. 1.). It consists of four parts: the input layer $H_{in}$, the LSTM layer $H_l$, the DenseNet layer $H_d$, and the residual output layer $H_{out}$. Compared with RDNet, ECGDeDRDNet introduces an additional step, specifically the LSTM layer $H_s$, denoted as $L_t$, which operates in loops within ECGDeDRDNet to pass the information about noise from previous steps to subsequent ones. Similar to RDNet, the relatively clean image $S_t$ produced by ECGDeDRDNet can be described as follows:

$$L_t = H_l(H_{in}(S_{t-1}, inputs), L_{t-1}) \qquad (6)$$
$$S_t = H_{out}(H_d(H_l(H_{in}(S_{t-1}, inputs), L_{t-1})) \qquad (7)$$

Here, $L_{t-1}$ and $L_t$ represent the hidden information of noise computed by the LSTM model during the $(t-1)^{th}$ and $t^{th}$ iterations, respectively.

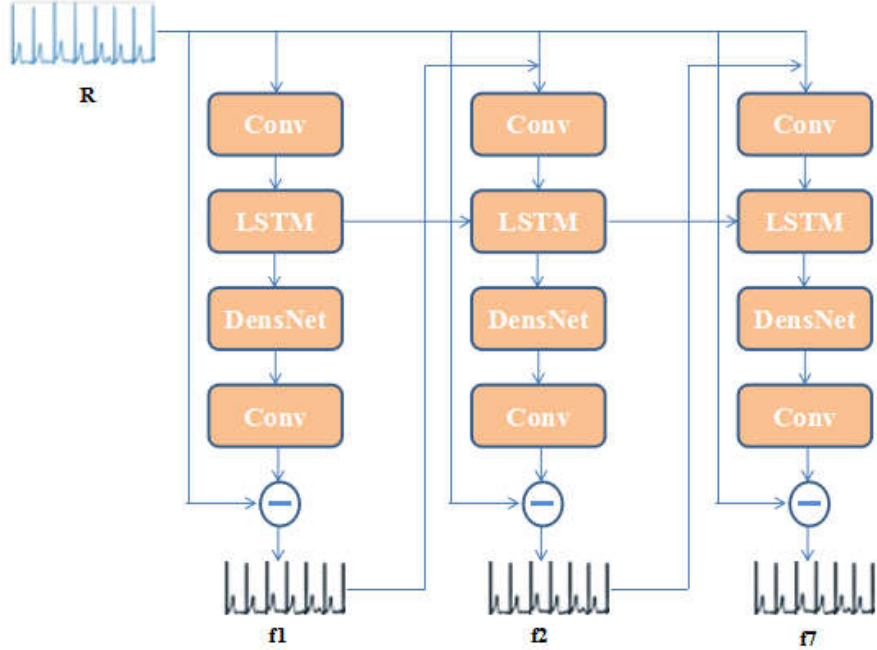

**Fig.3.** The framework of ECGDeDRDNet. $f_1$ denotes the output image of the noise removed from the first iteration, $f_7$ presents the output image of the 7$^{th}$ iteration.

As illustrated in Fig.3., the recurrent structure can be considered as multiple copies of the same structure, consisting of LSTM and DenseNet block, connected in series. Unlike a traditional recurrent framework, there is no recurrence within the entire network, each structure simply passes the current information to its successor. However, ECGDeDRDNet introduces two distinct information flow channels to enhance performance:

- LSTM Channel: The first channel, introduced by the LSTM model, primarily transmits noise-related information to the next stage. It enables the network to effectively utilize historical noise patterns.
- Recurrent Structure Channel: The second channel, facilitated by the recurrent structure, transfers the outputs after noise removal from the previous stage to the subsequent stage. It ensures that cleaned information is progressively refined throughout the network..

The two channels complement each other- one focuses on noise information, while the other handles relatively clean image data. This dual-channel approach enhances the overall performance of single image denoising by integrating both types of information effectively.

### 3.3 Evaluation Metrics and Loss Function

In this section, we first present the evaluation metrics used in our experiments and then describe the loss function in detail.

- **SSIM**

SSIM (Structural Similarity Index Measure, SSIM) [52] is widely employed to measure the structural similarity between two images. The corresponding mathematic expression is described as follows:

$$\text{SSIM}(S, H(x)) = \frac{(2\mu_S\mu_{H(x)}+c_1)(\sigma_{SH(x)}+c_2)}{(\mu_S^2+\mu_{H(x)}^2+c_1)(\sigma_S^2+\sigma_{H(x)}^2+c_2)} \quad (8)$$

Where $\mu_S$ represents the pixel mean value of the clean ECG image, $\mu_{H(x)}$ represents the pixel mean value of the denoised ECG image, $\sigma_{SH(x)}$ is the covariance of the clean ECG image S and the denoised image H(x). $\sigma_{SH}$, $\sigma_{H(x)}$ are the standard deviations of S and H(x), respectively. $c_1 = (k_1 * L)^2$ and $c_2 = (k_2 * L)^2$, are constants, where L denotes the dynamic range of pixel value. Typically, L = 255, $c_1 = 0.01$, and $c_2 = 0.03$.

Generally, the larger the SSIM value, the better the visual effect of the denoised images. If SSIM = 1, it indicates that the denoised image and the clean image are identical. Therefore, maximizing SSIM is often converted into a minimization problem using the following expression:

$$L_{SSIM} = 1 - SSIM(S, H(x)) \quad (9)$$

- **PSNR**

PSNR (Peak Single to Noise Ratio, PSNR) [53] represents the ratio of the maximum possible power of a signal and the destructive noise power that affects its representation accuracy. PSNR mainly compares the difference in the values of each pixel of the two images, which is typically referred to as "noise". If the two images are exactly identical, then the noise is 0 and the PSNR value is infinity. Conversely, if the two images differ significantly, the noise will be high and the PSNR will decrease accordingly. Therefore, the higher the PSNR value, the better the image quality. PSNR is defined as follows:

$$\text{PSNR} = 20\log_{10}\left(\frac{Peak}{\sqrt{MSE}}\right) \quad (10)$$

Where $Peak$ denotes the maximum value of the image pixel intensity. For 8 bit images, $Peak = 255$. MSE (Mean Squared Error, MSE) is calculated as $\text{MSE} = \frac{1}{N}\sum_{n=1}^{N}(H(x) - S)$.

- **L1-norm**

Let a training set $(S_i, X_i)$, $i = 1, 2, \ldots, n$, $n$ is the number of the training samples, $X_i$ is the noisy image and $S_i$ is the corresponding

clean image for $X_i$. L1-norm [54] is expressed as follows:

$$L_1 = \frac{1}{n}\sum_{i=1}^{n} |H(x)_i - S_i| \qquad (11)$$

L1-norm represents the sum of the absolute differences between the noisy images and clean images. It tends to shrink some feature weights towards 0.

- **L2-norm**

Different from L1-norm, L2-norm [55] is widely to penalize the large errors and calculate the least error between noisy images and clean images. The corresponding expression is described as follows:

$$L_2 = \frac{1}{n}\sum_{i=1}^{n} (H(x)_i - S_i)^2 \qquad (12)$$

- **Combined Loss Function**

To achieve optimal denoising results, based on the analysis of the relative metrics and extensive experiments, we use the following linear summation as the loss function in our study.

$$L_{com} = \beta * L_{SSIM} + (1 - \beta) * (L_1 + L_2) \qquad (13)$$

In our study, $\beta$ is set to 0.85.

## 4. Results and Discussion

4.1 Dataset

The MIT-BIH Arrhythmia Database [56] and the MIT-BIH Noise Stress Test Database [57] serve as experimental datasets in this study. The data from the MIT-BIH Arrhythmia Database is derived from clinical medical of Boston's Beth Israel Hospital. It includes a total of 48 records, each comprising signals from a modified limb lead II (MLII) and lead V1 (occasionally V2 or V5).

From the MIT-BIH Noise Stress Test Database, EM, BW and MA noise records are selected as noise data. These noise records were collected from volunteers using the professional ECG acquisition devices. By setting various SNRs, noise data is added to the original ECG signal to produce the training and test data for subsequent experiments. Given the periodic nature of ECG signals, single image segments are selected for the validating our proposed method. The MIT-BIH Arrhythmia Database contains approximately 92000 images, from which 13862 image segments are extracted by cutting the ECG signals at a rate of 5 seconds per segment, with each segment containing 1800 data points (equivalent to 5 seconds of signal).

In this study, the 13862 image fragments are divided into two parts: the training set and the testing set, the former accounts for 80% of the segments, and the latter accounts for 20% of the segments.

## 4.2 Additional Evaluation Metrics for ECG Denoising

In addition to PSNR, SSIM metrics, which are common used to assess image denoising methods, we introduce two additional metrics-the signal-noise ratio (SNR) and the root mean square error (RMSE) to compare with ECG denoising methods. These metrics are defined as follows:

$$\text{SNR} = 10lg\frac{\sum_{i=1}^{N}(S_i)^2}{\sum_{i=1}^{n}(H(x)_i - S_i)^2} \quad (14)$$

$$RMSE = \sqrt{\frac{1}{n}\sum_{i=1}^{n}(H(x)_i - S_i)^2} \quad (15)$$

Where $S_i$ represents the clean signal at sample point i, $H(x)_i$ represents the denoised signal at sample point i, N is the total number of samples. The SNR represents the ratio of the signal power to the noise power contained in the signal. A larger SNR value indicates a better noise removal effect. The RMSE measures the differences between the denoised signal and the clean signal, implying better performance of the denoising method.

As described in Eq.(14)-(15), SNR quantifies the ratio of the signal power to the noise power. The larger the SNR value, the more effective the noise removal. RMSE describes the difference between the denoised signal and the clean signal. The smaller the RMSE value, the closer the denoised signal is to the clean signal, indicating better denoising performance.

## 4.3 Experiment setup

In our experiments, a large number of training image patches, each with a size of 100*100 pixels, are cropped from the training images. Our proposed model is implemented on the Pytorch platform, utilizing Leaky ReLU [58] for nonlinear operations. The model trainings and all subsequent experiments were conducted on a PC equipped with Intel Core i5-13600kf CPU and an NVIDA GeFore RTX 3090GPU, running Windows 11. For each comparison approach, we followed the parameter settings recommended by the original authors to ensure optimal performance. These parameter configurations are designed to achieve the best results for each considered method. The specific parameter settings used for each method are summarized in Table 1.

**Table 1** The specific parameter settings

| Parameter | size | Parameter name | size |
|---|---|---|---|

| name | | | |
|---|---|---|---|
| Image pathces | 100*100 | Stage of ECGDeDRDNet (t) | 7 |
| Bathes | 5 | Learning rate | 0.0001 |

### 4.4 Effects of Parameter $L_1, L_2, L_{SSIM}, L_{com}$

Table 2 provides the performance results of our proposed ECGDeDRDNet method evaluated using three different metrics PNSR, SSIM and the loss function $L_1, L_2, L_{SSIM}$. These results were obtained after running the model for 100 iterations. From Table 2, it is evident that $L_{SSIM}$ outperforms the other two metrics of $L_1$ and $L_2$ in terms of both PSNR and SSIM. Specifically, $L_{SSIM}$ achieves a PSNR value of 88.07 and an SSIM value of 0.847. Surprisingly, the combined metric of $L_{com}$ achieves even better results than the individual metric. it achieves the highest PSNR value of 88.39 and an SSIM of 0.847. Given these findings, $L_{com}$ is selected for conducting further experiments on ECG denoising.

**Table 2** Performance Results of ECGDeDRDNet method evaluated using three different metrics

| Metric | PSNR | SSIM |
|---|---|---|
| $L_1$-norm | 87.88 | 0.841 |
| $L_2$-norm | 87.10 | 0.837 |
| $L_{SSIM}$ | 88.07 | **0.847** |
| $L_{com}$ | **88.39** | **0.847** |

*The highest value in each row is in bold.*

### 4.5 Comparison with other denoising methods
#### 4.5.1 Image denoising performance

To highlight the image denoising performance of ECGDeDRDNet, we compare it with RDNet in terms of PSNR, SSIM. From Table 3, it is evident that our proposed method of ECGDeDRDNet achieved the highest value for both metrics. Our proposed method demonstrates a PSNR improvement of 1.3% and an SSIM improvement of 2.7% compared to RDNet. It indicates that ECGDeDRDNet shows superior denoising results in terms of image noise reduction.

**Table 3** Performance Comparison with RDNet

| Method | PSNR | SSIM |
|---|---|---|
| RDNet | 87.28 | 0.824 |
| ECGDeDRDNet | **88.39** | **0.847** |

*The highest value in each row is in bold.*

4.5.2 ECG denoising performance

In this study, the proposed method, ECGDeDRDNet, is primarily employed for denoising ECG signal. Here, we focus on the comparison ECGDeDRDNet with existing ECG denosing methods in terms of SNR and *RMSE*.

We selected ten records from the MIT-BIH Arrhythmia Database (numbers 103, 105, 111, 116, 122, 205, 213, 219, 223, 230) as experimental data, and added EM, BW and MA noise signals chosen from the MIT-BIH Noise Stress Test Database to each record at SNR levels of 0dB, 1.25dB and 5dB. Additionally, we considered the mixed noise cases with multiple noise types (EM+BW, EM+MA, BW+MA, and EM+BW+MA), where these noise types were mixed into the original clean signal in equal proportions.

This experiment compares the noise reduction performance of S-Transform [59], improved WT [60], and SDAE [61] and our proposed method of ECGDeDRDNet. Tables 4-6 show the results for ten records with different noise types (MA noise, the EM noise and the BW noise). It is observed that in all cases, our proposed method achieves the highest SNR values, both individuals and on average. The lowest RMSE values are also achieved on average, though some individual records show better performance by other methods such as SDAE. For example, when BW noise signals are added to each record with an SNR of 5dB, except for the records NO.205, 213, 230, our proposed method achieves the lowest RMSE value. In other cases, the lowest RMSE values are obtained by SDAE. Therefore, our proposed method outperforms SADE slightly on RMSE. This could be attributed to the fact that SDAE mainly focuses on signal denoising, while our method can handle both image and signal denoising and may require further optimization for signal denoising.

To highlight the denoising effectiveness, we take record number 103 as example, with an SNR of 1.25dB. We present the noise reduction effect diagrams of single ECGDeDRDNet model for different types of noise (EM noise, BW noise and MA noise) in Fig.4. In subfigures, the upper line represents the noisy signal, the middle line represents the denoised signal and the bottom line represents the clean signal. From the dinoised signal (middle line), it is clear that the entire denoised ECG signals are very close to the clean ECG signals, with only subtle waveform differences. Surprisingly, the noise reduction of $2^{th}$ of ECG signals result in a more prominent R-wave compared to $1^{th}$ series, providing a clearer basis for doctors' clinical diagnosis. These results suggest that our method is capable of effectively reducing EM noise, the BW noise and the MA noise in ECG signals using a single ECGDeDRDNet model.

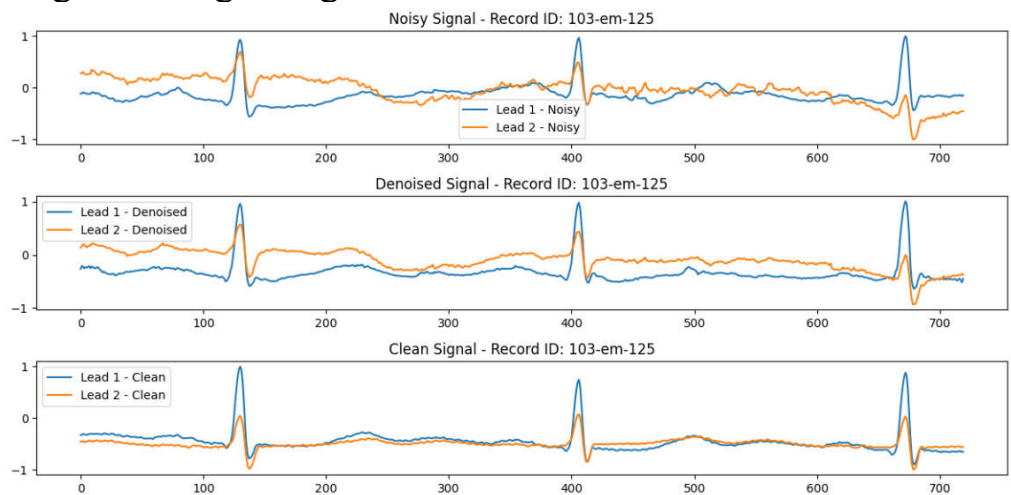

(a) ECG signal results with EM noise

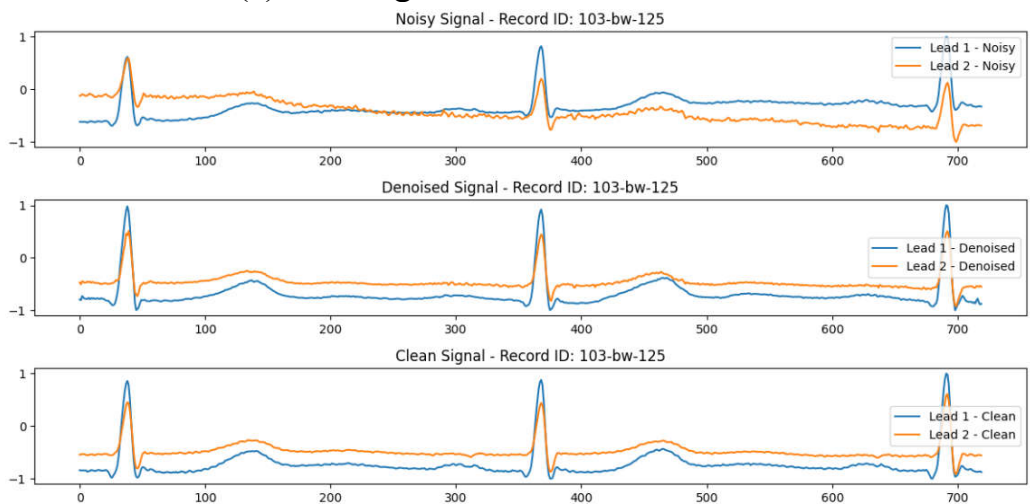

(b) ECG signal results with BW noise

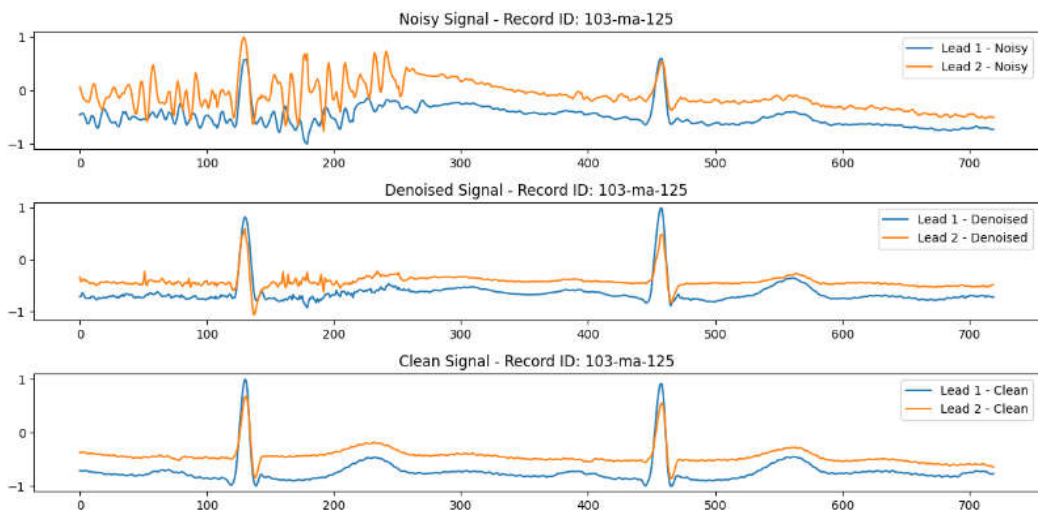

(c) ECG signal results with MA noise

**Fig.4.** ECG signal denoised results with EM, BW, MA noise

Similarly, as depicted in Fig.5, the results of ECG denosing for removing the mixed noise are presented. From subfigures (a) to (d) represent the denoised results for the following mixed noise types: EM+BW, EM+MA, BW+MA, and EM+BW+MA. In each subfigure, the upper line represents the noisy signal, the middle line represents the denoised signal, and the bottom line represents the clean signal. It is observed that the overall denoised ECG signals are very close to the clean ECG signals, similar to what was shown in Fig.4. This indicates that our proposed method is effectiveness in reducing mixed noise in ECG signals using a single ECGDeDRDNet model. The findings suggest that our proposed method is highly effective reducing various types of mixed noise in ECG signals.

Specifically, when mixed noise EM+BW is tested, the denoised signals closely match the clean signals, demonstrating the effectiveness in handling this combination of noise. Similarly, when mixed noise EM+MA is experimented, the denoised signals show minimal difference from the clean signals. Interestingly, the denoised signals generated BW+MA remain close to the clean signals, indicating robust performance even with multiple noise types. Despite the complexity of this mixed noise scenario (EM+BW+MA), the denoised signals still exhibit high fidelity to the clean signals.

These results highlight the versatility and effectiveness of ECGDeDRDNet in denoising ECG signals under different noise conditions. Moreover, the removal of mixed noise by ECGDeDRDNet provides valuable insights and a potential denoising strategy for future remote ECG monitoring systems.

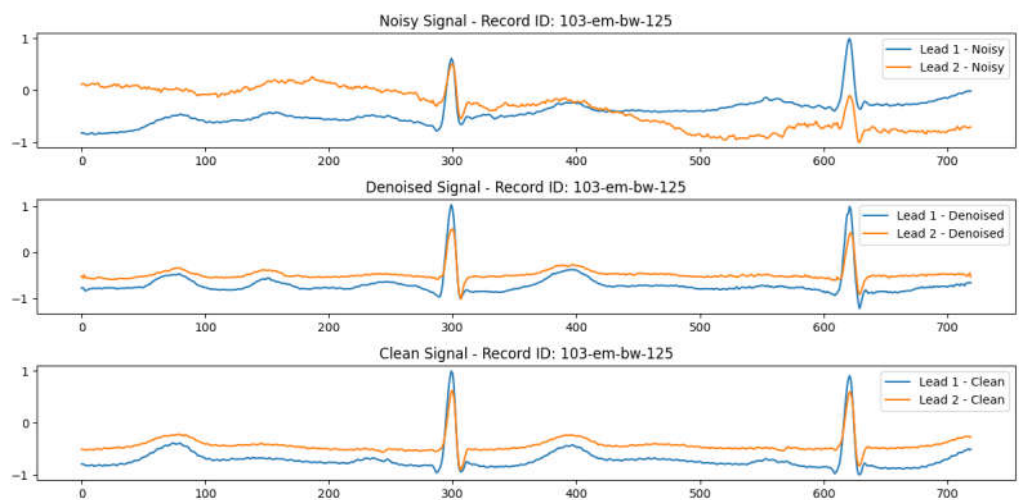

(a) ECG signal denoised results with noise of EM+BW

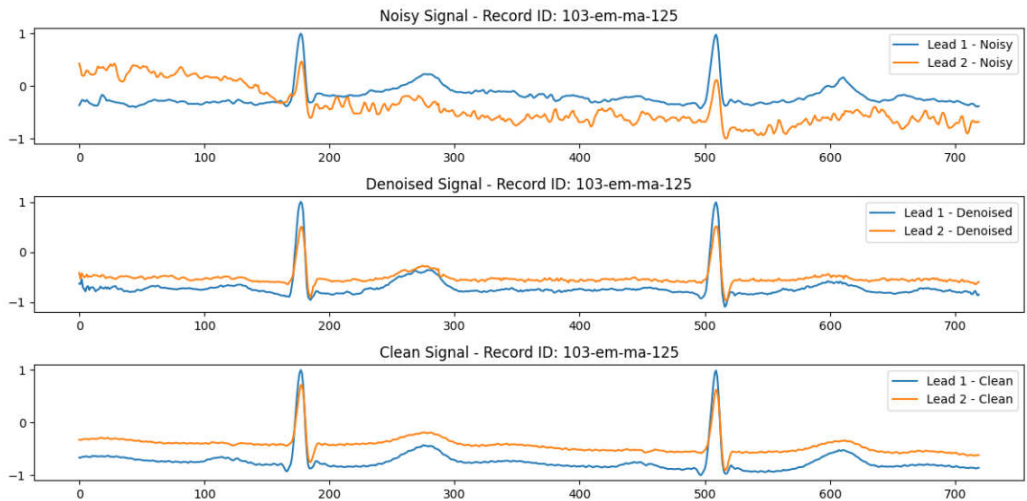

(b) ECG signal denoised results with noise of EM+MA

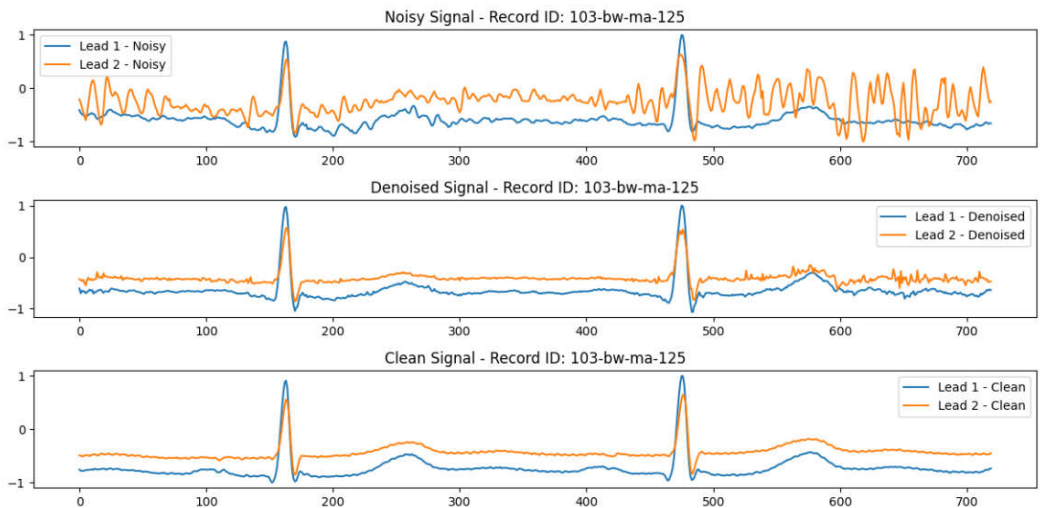

(c) ECG signal denoised results with noise of BW+MA

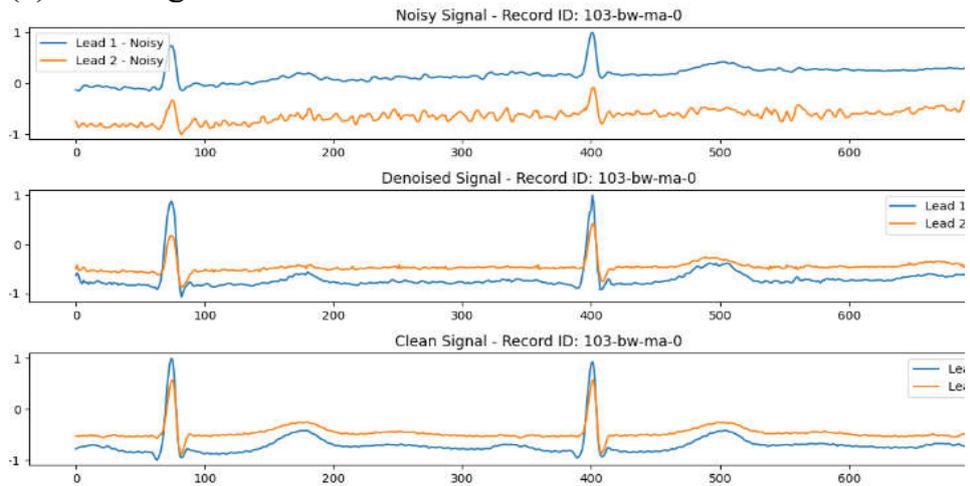

(d) ECG signal denoised results with noise of EM+BW+MA
**Fig.5.** ECG signal denoised results with mixed noise

**Table 4** Experimental results for removing EM noise

| | MIT-BIH data NO. | | 103 | 105 | 111 | 116 | 122 | 205 | 213 | 219 | 223 | 230 | Average |
|---|---|---|---|---|---|---|---|---|---|---|---|---|---|
| S-Transform | 0dB | SNR | 6.41 | 6.13 | 5.45 | 5.47 | 5.87 | 5.59 | 5.85 | 6.05 | 6.21 | 6.29 | 5.932 |
| | | RMSE | 0.478 | 0.493 | 0.534 | 0.533 | 0.509 | 0.525 | 0.51 | 0.498 | 0.489 | 0.484 | 0.505 |
| | 1.25dB | SNR | 7.47 | 7.35 | 6.4 | 6.32 | 6.96 | 6.47 | 7.06 | 7.17 | 7.45 | 7.48 | 7.013 |
| | | RMSE | 0.423 | 0.429 | 0.478 | 0.483 | 0.449 | 0.475 | 0.444 | 0.438 | 0.424 | 0.423 | 0.447 |
| | 5dB | SNR | 10.32 | 10.4 | 8.54 | 8.32 | 9.6 | 8.55 | 10.12 | 10.04 | 10.74 | 10.45 | 9.708 |
| | | RMSE | 0.305 | 0.302 | 0.374 | 0.384 | 0.331 | 0.374 | 0.312 | 0.315 | 0.29 | 0.3 | 0.329 |
| WT | 0dB | SNR | 9.51 | 21.47 | 9.34 | 9.63 | 0.71 | 18.29 | 15.02 | 13.15 | 18.11 | 11.99 | 13.62 |
| | | RMSE | 0.138 | 0.044 | 0.171 | 0.138 | 0.128 | 0.063 | 0.067 | 0.076 | 0.058 | 0.132 | 0.102 |
| | 1.25dB | SNR | 10.37 | 23.08 | 9.65 | 10.56 | 9.51 | 20.3 | 15.8 | 13.85 | 21.1 | 12.93 | 14.72 |
| | | RMSE | 0.125 | 0.037 | 0.165 | 0.124 | 0.132 | 0.05 | 0.061 | 0.07 | 0.04 | 0.119 | 0.092 |
| | 5dB | SNR | 13.07 | 28.37 | 14.96 | 13.77 | 8.65 | 21.36 | 19.2 | 16.2 | 24.05 | 15.91 | 17.55 |
| | | RMSE | 0.091 | **0.02** | 0.09 | 0.086 | 0.145 | 0.043 | 0.04 | 0.054 | 0.029 | 0.084 | 0.068 |
| SDAE | 0dB | SNR | 18.94 | 23.45 | 22.33 | 19.18 | 17.87 | 20.08 | 19.2 | 17.53 | 22.66 | 20.79 | 20.2 |
| | | RMSE | 0.047 | 0.035 | 0.038 | 0.046 | 0.047 | 0.05 | 0.039 | 0.045 | 0.033 | 48 | 0.043 |
| | 1.25dB | SNR | 19.07 | 23.82 | 22.43 | 19.69 | 18.98 | 20.11 | 19.74 | 18.32 | 23.2 | 20.91 | 20.63 |
| | | RMSE | 0.046 | 0.033 | 0.038 | 0.043 | 0.042 | 0.049 | 0.037 | 0.041 | 0.031 | 0.047 | 0.041 |
| | 5dB | SNR | 19.3 | 24.56 | 22.66 | 21 | 21.16 | 20.24 | 20.98 | 20.08 | 24.39 | 21.4 | 21.58 |
| | | RMSE | 0.041 | 0.03 | 0.037 | 0.037 | 0.033 | 0.049 | 0.033 | 0.034 | 0.027 | 0.044 | 0.037 |

| | | | | | | | | | | | | |
|---|---|---|---|---|---|---|---|---|---|---|---|---|
| ECGDeDRDNet | 0db | SNR | **24.406** | **21.26** | **16.91** | **22.297** | **23.451** | **25.687** | **18.279** | **19.237** | **19.006** | **19.635** | **21.017** |
| | | RMSE | 0.022 | 0.052 | 0.061 | 0.03 | 0.026 | 0.021 | 0.006 | 0.054 | 0.031 | 0.027 | **0.033** |
| | 1.25db | SNR | **25.096** | **21.896** | **17.527** | **22.981** | **24.227** | **26.257** | **18.866** | **19.766** | **19.66** | **21.187** | **21.746** |
| | | RMSE | 0.013 | 0.041 | 0.05 | **0.02** | **0.014** | 0.045 | 0.039 | 0.045 | 0.022 | 0.042 | **0.033** |
| | 5db | SNR | **26.736** | **23.231** | **19.112** | **24.295** | **25.424** | **27.315** | **19.414** | **21.032** | **22.174** | **22.529** | **23.126** |
| | | RMSE | 0.043 | 0.022 | 0.027 | 0.024 | 0.03 | 0.015 | 0.013 | 0.026 | 0.013 | 0.011 | **0.022** |

Table 5 Experimental results for removing BW noise

| | MIT-BIH data NO. | | 103 | 105 | 111 | 116 | 122 | 205 | 213 | 219 | 223 | 230 | Average |
|---|---|---|---|---|---|---|---|---|---|---|---|---|---|
| S-Transform | 0dB | SNR | 11.400 | 11.560 | 9.220 | 9.010 | 10.580 | 9.310 | 11.570 | 11.550 | 12.140 | 11.680 | 10.802 |
| | | RMSE | 0.269 | 0.264 | 0.346 | 0.354 | 0.296 | 0.342 | 0.264 | 0.265 | 0.247 | 0.261 | 0.291 |
| | 1.25dB | SNR | 12.060 | 12.230 | 9.610 | 9.350 | 11.170 | 9.700 | 12.220 | 12.220 | 12.950 | 12.350 | 11.386 |
| | | RMSE | 0.249 | 0.245 | 0.331 | 0.341 | 0.276 | 0.327 | 0.245 | 0.245 | 0.225 | 0.241 | 0.273 |
| | 5dB | SNR | 13.540 | 13.770 | 10.410 | 10.040 | 12.380 | 10.460 | 13.770 | 13.650 | 14.810 | 13.890 | 12.672 |
| | | RMSE | 0.210 | 0.205 | 0.302 | 0.315 | 0.240 | 0.300 | 0.205 | 0.208 | 0.182 | 0.202 | 0.237 |
| WT | 0dB | SNR | 14.870 | 31.53 | 18.41 | 20 | 9. 12 | 22.64 | 20.83 | 18.69 | 17.34 | 22.23 | 19.57 |
| | | RMSE | 0.074 | 0.014 | 0.06 | 0.042 | 0. 138 | 0.037 | 0.034 | 0.04 | 0.063 | 0.041 | 0.054 |
| | 1.25dB | SNR | 14.88 | 31.91 | 18.42 | 20.06 | 8.54 | 22.73 | 20.47 | 20.22 | 17.4 | 22.22 | 19.69 |

|  |  |  | 103 | 105 | 111 | 116 | 122 | 205 | 213 | 219 | 223 | 230 | Average |
|---|---|---|---|---|---|---|---|---|---|---|---|---|---|
|  |  | RMSE | 0.074 | 0.013 | 0.06 | 0.042 | 0.147 | 0.037 | 0.036 | 0.034 | 0.063 | 0.041 | 0.055 |
|  | 5dB | SNR | 14.9 | 32.71 | 18.43 | 20.1 | 8.22 | 22.91 | 19.11 | 21.44 | 17.51 | 22.18 | 19.75 |
|  |  | RMSE | 0.074 | 0.012 | 0.06 | 0.041 | 0.153 | 0.036 | 0.042 | 0.029 | 0.062 | 0.041 | 0.055 |
|  | 0dB | SNR | 20.38 | 24.9 | 23.04 | 18.84 | 19.48 | 20.08 | 19.92 | 19.3 | 22.94 | 20.53 | 20.94 |
|  |  | RMSE | 0.038 | 0.029 | 0.035 | 0.047 | 0.04 | 0.05 | 0.036 | 0.037 | 0.031 | 0.049 | 0.039 |
| SDAE | 1.25dB | SNR | 20.55 | 23.27 | 23.07 | 19.53 | 19.9 | 20.12 | 20.32 | 19.83 | 23.74 | 20.67 | 21.3 |
|  |  | RMSE | 0.038 | 0.028 | 0.035 | 0.043 | 0.038 | 0.049 | 0.035 | 0.035 | 0.029 | 0.048 | 0.038 |
|  | 5dB | SNR | 20.77 | 25.47 | 23.03 | 21.6 | 21 | 20.3 | 21.34 | 21.15 | 25.41 | 21.03 | 22.11 |
|  |  | RMSE | **0.037** | **0.027** | **0.035** | **0.034** | **0.034** | 0.048 | 0.031 | 0.03 | **0.024** | 0.046 | 0 035 |
|  | 0db | SNR | **27.118** | **25.238** | **19.141** | **21.091** | **20.906** | **27.483** | **21.049** | **21.312** | **22.810** | **22.991** | **22.914** |
|  |  | RMSE | 0.046 | 0.036 | 0.022 | 0.048 | 0.046 | 0.035 | 0.029 | 0.036 | 0.039 | 0.039 | 0.038 |
| ECGDeDRDNet | 1.25db | SNR | **27.553** | **23.702** | **19.646** | **25.428** | **26.227** | **27.778** | **17.496** | **17.737** | **19.250** | **19.367** | **22.418** |
|  |  | RMSE | 0.046 | 0.036 | 0.022 | 0.042 | 0.046 | 0.035 | 0.029 | 0.036 | 0.039 | 0.039 | 0.037 |
|  | 5db | SNR | **26.697** | **20.802** | **18.980** | **24.187** | **25.002** | **26.417** | **20.673** | **20.774** | **22.307** | **22.317** | **22.815** |
|  |  | RMSE | 0.049 | 0.036 | 0.037 | 0.038 | 0.035 | 0.026 | 0.026 | 0.028 | 0.034 | 0.030 | 0.034 |

**Table 6** Experimental results for removing MA noise

| MIT-BIH data NO. |  |  | 103 | 105 | 111 | 116 | 122 | 205 | 213 | 219 | 223 | 230 | Average |
|---|---|---|---|---|---|---|---|---|---|---|---|---|---|
| S-Transform | 0dB | SNR | 10.41 | 10.02 | 8.21 | 8.19 | 9.2 | 8.32 | 8.79 | 10.05 | 9.95 | 8.7 | 9.184 |
|  |  | RMSE | 0.302 | 0.316 | 0.389 | 0.389 | 0.347 | 0.384 | 0.363 | 0.314 | 0.318 | 0.367 | 0.3489 |
|  | 1.25dB | SNR | 10.89 | 10.42 | 8.66 | 8.51 | 9.67 | 8.61 | 9.67 | 10.61 | 10.56 | 9.01 | 9.661 |
|  |  | RMSE | 0.286 | 0.301 | 0.369 | 0.375 | 0.328 | 0.371 | 0.329 | 0.295 | 0.297 | 0.355 | 0.3306 |
|  | 5dB | SNR | 12 63 | 12 76 | 9 94 | 9 75 | 11 69 | 9 91 | 12 53 | 12 89 | 13 44 | 10 15 | 11 569 |
|  |  | RMSE | 0 234 | 0 23 | 0 318 | 0 325 | 0 26 | 0 32 | 0 236 | 0 227 | 0 213 | 0 311 | 0 2674 |
|  | 0dB | SNR | 19.66 | 22.09 | 20.02 | 12.44 | 6.71 | 21.23 | 11.83 | 7.33 | 18.58 | 18.03 | 15.79 |
|  |  | RMSE | 0.044 | 0.040 | 0.050 | 0. 100 | 0. 182 | 0.044 | 0.096 | 0. 149 | 0.055 | 0.066 | 0.083 |
| WT | 1.25dB | SNR | 16.87 | 22.49 | 19.69 | 14.40 | 7.43 | 20.24 | 13.27 | 8.68 | 20.32 | 18.86 | 16.23 |
|  |  | RMSE | 0.060 | 0.039 | 0.052 | 0.080 | 0. 167 | 0.049 | 0.081 | 0. 127 | 0.045 | 0.060 | 0.076 |
|  | 5dB | SNR | 15.79 | 24. 11 | 18.81 | 19. 15 | 11. 14 | 16.51 | 18.94 | 14.55 | 21.41 | 21. 14 | 18. 16 |
|  |  | RMSE | 0.067 | 0.032 | 0.057 | 0.046 | 0. 109 | 0.075 | 0.042 | 0.065 | 0.040 | 0.046 | 0.058 |
|  | 0dB | SNR | 18 92 | 22 97 | 22 90 | 17 92 | 17 96 | 20 03 | 18 18 | 16 18 | 20 29 | 21 11 | 19 65 |
|  |  | RMSE | 0 046 | 0 036 | 0 036 | 0 053 | 0 047 | 0 050 | 0 044 | 0 049 | 0 043 | 0 046 | 0 045 |
| SDAE | 1.25dB | SNR | 19 07 | 23 30 | 22 88 | 18 48 | 18 09 | 20 03 | 18 70 | 17 66 | 21 21 | 21 22 | 20 06 |
|  |  | RMSE | 0.045 | 0.035 | 0.036 | 0.049 | 0.042 | 0.050 | 0.041 | 0.044 | 0.039 | 0.045 | 0.043 |
|  | 5dB | SNR | 19.31 | 24. 12 | 22.95 | 20.61 | 21.73 | 20. 14 | 20. 15 | 20.05 | 23.69 | 21.33 | 21.41 |
|  |  | RMSE | 0.044 | 0.032 | 0.035 | 0.038 | 0.031 | 0.050 | 0.036 | 0.033 | 0.029 | 0.044 | 0.037 |
|  | 0db | SNR | **22.198** | **19.638** | **16.977** | **20.135** | **21.184** | **22.594** | **17.710** | **18.783** | **18.698** | **18.809** | **19.673** |
| ECGDeDRDNet |  | RMSE | 0.030 | 0.050 | 0.055 | 0.035 | 0.032 | 0.019 | 0.003 | 0.059 | 0.032 | 0.038 | 0.035 |
|  | 1.25db | SNR | **23.971** | **21.279** | **17.204** | **21.798** | **22.824** | **24.344** | **18.371** | **18.688** | **19.704** | **20.402** | **20.858** |

|  |      | 0db | 0db | 0db | 0db | 0db | 0db | 0db | 0db | 0db | 0db | 0db |
| --- | --- | --- | --- | --- | --- | --- | --- | --- | --- | --- | --- | --- |
|  | RMSE | 0.018 | 0.039 | 0.044 | 0.025 | 0.022 | 0.008 | 0.002 | 0.049 | 0.122 | 0.082 | 0.041 |
| 5db | SNR | **24.939** | **21.726** | **17.901** | **22.313** | **23.301** | **24.831** | **19.134** | **19.243** | **20.452** | **20.987** | **21.483** |
|  | RMSE | 0.044 | 0.048 | 0.020 | 0.047 | 0.042 | 0.040 | 0.038 | 0.026 | 0.031 | 0.029 | 0.036 |

## 5. Conclusion

In this study, we have investigated a deep learning-based method for ECG removal method using a Double Recurrent Dense Network, named as ECGDeDRDNet. Our approach involved several key steps:

- Model Development: We developed a recurrent dense network model to leverage the strengths of both recurrent structures and dense connections.
- LSTM Integration: To enhance performance in single image noise reduction, we incorporated Long Short-Term Memory (LSTM) units into our iterative process.
- Loss Function Innovation: We proposed a new loss function that serves as a more effective alternative for achieving high-quality signal denoising.
- Validation: The quality of noise removal was validated from two perspectives:
  I) Image denoising, evaluated using PSNR and SSIM metrics.
  II) Signal denoising, evaluated using SNR and RMSE metrics.

The experimental results on the MIT-BIH database demonstrate the effectiveness of our proposed method. Notably, the removal of mixed noise highlights the potential of ECGDeDRDNet for future remote ECG monitoring systems.

Despite achieving satisfactory experimental results in both image and signal denoising, the advantages in signal denoising are not as pronounced as those in image denoising. To address this and further improve the performance of ECG signal denoising, we propose the following future work directions:

- Optimization of Deep Denoising Models: Enhance the existing deep denoising models to better handle the unique

characteristics of ECG signals.
- Mitigation of Deep Learning Limitations: Adopt reasonable methods to overcome common challenges in deep learning, such as overfitting and the need for interpretable mechanisms. This includes developing strategies to ensure robustness and generalizability of the models.

By focusing on above areas, we aim to refine our approach and achieve even better performance in ECG signal denoising, ultimately contributing to more reliable and accurate remote ECG monitoring systems.

## Acknowledgment

The work was supported by National Natural Science Foundation of China (61672217, 61932010), the Natural Science Foundation of Hunan Provincial (2023JJ30112), and the Key Project of the Education Department of Hunan Province (24A0568).

## References

[1] R. Dey, P. K. Samanta, R. P. Chokda, B. P. De, B. Appasani, A. Srinivasulu, and N. Philibert, "Graphene-based electrodes for ECG signal monitoring: Fabrication methodologies, challenges and future directions," *Cogent Engineering,* vol. 10, Dec 31 2023.

[2] M. Kolhar and A. M. Al Rajeh, "Deep learning hybrid model ECG classification using AlexNet and parallel dual branch fusion network model," *Scientific Reports,* vol. 14, Nov 6 2024.

[3] T. Li, N. Wang, H. X. Yang, M. F. Dou, and X. W. Yang, "A lead-aware hierarchical convolutional neural network for arrhythmia detection in electrocardiogram," *Signal Image and Video Processing,* vol. 19, Jan 2025.

[4] G. Warminski, L. Kalinczuk, M. Orczykowski, P. Urbanek, R. Bodalski, K. Zielinski, M. Gandor, F. Palka, M. Jaworski, G. S. Mintz, I. Kowalik, A. Hasiec, M. Bilinska, P. Plawiak, and L. Szumowski, "Artificial intelligence analysis of ECG signals to predict arrhythmia recurrence after cryoballoon ablation of atrial fibrillation," *European Heart Journal,* vol. 43, pp. 559-559, Oct 2022.

[5] L. Li, J. Camps, Z. Wang, M. Beetz, A. Banerjee, B. Rodriguez, and V. Grau, "Toward Enabling Cardiac Digital Twins of Myocardial Infarction Using Deep Computational Models for Inverse Inference," *Ieee Transactions on Medical Imaging,* vol. 43, pp. 2466-2478, Jul 2024.

[6] H. Y. Lee, Y. J. Kim, K. H. Lee, J. H. Lee, S. P. Cho, J. Park, I. Park, and H. Youk, "Substantiation and Effectiveness of Remote Monitoring System Based on IoMT Using Portable ECG Device," *Bioengineering-Basel,* vol. 11, Aug 2024.

[7] W. Wei, L. X. Lu, Y. Hao, S. Kang, Y. H. Liu, J. Yu, W. L. Chen, and C. H. Fan, "Application of remote electrocardiogram monitoring systems in chest

pain centers for patients with high-risk chest pain," *Technology and Health Care,* vol. 32, pp. 411-421, 2024.
[8] S. S. Ahmed, T. Ahmed, E. G. Abdalla, A. A. M. Humidan, A. M. A. Daffalla, A. T. Elgabani, M. A. Abdelrahem, T. Bilal, and A. A. Ibrahim, "Preoperative ECG Abnormalities Among Patients Who Underwent Elective Surgical Operations at the Kuwaiti Specialised Hospital, Khartoum, Sudan: A Cross-Sectional Study," *CUREUS JOURNAL OF MEDICAL SCIENCE,* vol. 16, Feb 24 2024.
[9] P. Bhattarai and M. Karki, "Role of ECG in the Accidental Finding of an Atrioventricular Septal Defect in an Asymptomatic Patient Undergoing Cosmetic Surgery," *CUREUS JOURNAL OF MEDICAL SCIENCE,* vol. 16, Jan 16 2024.
[10] M. Molinari, S. Setti, N. D. Brunetti, N. Di Nunno, M. A. Cattabiani, and G. Molinari, "Different patterns of pre-excitation in a large Italian cohort of asymptomatic non-competitive athletes evaluated by telecardiology screening: Prevalence and ECG features," *Ijc Heart & Vasculature,* vol. 55, Dec 2024.
[11] A. Callanan, D. Quinlan, P. M. Kearney, S. O'Sullivan, G. Zhi, A. Crichton, M. W. Howell, C. Bradley, and C. Buckley, "Opportunistic atrial fibrillation screening in primary care in Ireland: results of a pilot screening programme," *Open Heart,* vol. 11, May 2024.
[12] M. Khalili, H. GholamHosseini, A. Lowe, and M. M. Y. Kuo, "Motion artifacts in capacitive ECG monitoring systems: a review of existing models and reduction techniques," *Medical & Biological Engineering & Computing,* vol. 62, pp. 3599-3622, Dec 2024.
[13] A. Joutsen, A. Cömert, E. Kaappa, K. Vanhatalo, J. Riistama, A. Vehkaoja, and H. Eskola, "ECG signal quality in intermittent long-term dry electrode recordings with controlled motion artifacts," *Scientific Reports,* vol. 14, Apr 17 2024.
[14] H. Y. Li, G. Ditzler, J. Roveda, and A. Li, "DeScoD-ECG: Deep Score-Based Diffusion Model for ECG Baseline Wander and Noise Removal," *Ieee Journal of Biomedical and Health Informatics,* vol. 28, pp. 5081-5091, Sep 2024.
[15] N. Ayman, A. Hesham, R. Ehab, M. Emad, I. M. El-Badawy, and Z. Omar, "ECG Baseline Wander Reduction Using Empirical Mode Decomposition: A Case Study on Preterm Infants With Bradycardia," *2024 Ieee 8th International Conference on Signal and Image Processing Applications, Icsipa,* 2024.
[16] J. V. Vadakkan, M. S. Manikandan, and L. R. Cenkeramaddi, "Lightweight Time-Domain Statistical Parameter Based Muscle Artifacts Detection for Trustworthy Wearable ECG Analysis Devices," *2024 Ieee 19th Conference on Industrial Electronics and Applications, Iciea 2024,* 2024.
[17] M. Chen, Y. J. Li, L. T. Zhang, L. Liu, B. K. Han, W. Z. Shi, and S. S. Wei, "Elimination of Random Mixed Noise in ECG Using Convolutional Denoising Autoencoder With Transformer Encoder," *Ieee Journal of Biomedical and Health Informatics,* vol. 28, pp. 1993-2004, Apr 2024.
[18] O. Mohguen, "Noise reduction and QRS detection in ECG signal using EEMD with modified sigmoid thresholding," *Biomedical Engineering-Biomedizinische Technik,* Sep 5 2023.
[19] S. H. Liu, C. H. Hsieh, W. Chen, and T. H. Tan, "ECG Noise Cancellation Based on Grey Spectral Noise Estimation," *Sensors,* vol. 19, 2019.
[20] M. Sraitih and Y. Jabrane, "A denoising performance comparison based on ECG Signal Decomposition and local means filtering," *Biomedical Signal Processing and Control,* vol. 69, Aug 2021.
[21] M. A. Awal, S. S. Mostafa, M. Ahmad, and M. A. Rashid, "An adaptive level dependent wavelet thresholding for ECG denoising," *Biocybernetics & Biomedical Engineering,* vol. 34, pp. 238-249, 2014.

[22] L. El Bouny, M. Khalil, and A. Adib, "A Wavelet Denoising and Teager Energy Operator-Based Method for Automatic QRS Complex Detection in ECG Signal," *Circuits Systems and Signal Processing,* vol. 39, pp. 4943-4979, Oct 2020.
[23] M. Talbi, "A New ECG Denoising Technique Based on LWT and TVM," *Circuits Systems and Signal Processing,* vol. 40, pp. 6284-6300, Dec 2021.
[24] C. P. Swamy, B. F. Besheli, L. R. F. Branco, N. R. Provenza, S. A. Sheth, W. K. Goodman, A. Viswanathan, and N. F. Ince, "Pulsation artifact removal from intra-operatively recorded local field potentials using sparse signal processing and data-specific dictionary," *2023 45th Annual International Conference of the Ieee Engineering in Medicine & Biology Society, Embc,* 2023.
[25] H. Shi, R. X. Liu, C. F. Chen, M. L. Shu, and Y. L. Wang, "ECG Baseline Estimation and Denoising With Group Sparse Regularization," *Ieee Access,* vol. 9, pp. 23595-23607, 2021.
[26] A. Dorostghol, A. Maghsoudpour, A. Ghaffari, and M. Nikkhah-Bahrami, "Power line interference and baseline wander removal from ECG signals using local characteristic decomposition," *Journal of Instrumentation,* vol. 17, Jun 2022.
[27] I. Christov, T. Neycheva, R. Schmid, T. Stoyanov, and R. Abächerli, "Pseudo-real-time low-pass filter in ECG, self-adjustable to the frequency spectra of the waves," *Medical & Biological Engineering & Computing,* vol. 55, pp. 1579-1588, Sep 2017.
[28] B. Y. Liu and Y. L. Li, "ECG signal denoising based on similar segments cooperative filtering," *Biomedical Signal Processing and Control,* vol. 68, Jul 2021.
[29] M. Chandra, P. Goel, A. Anand, and A. Kar, "Design and analysis of improved high-speed adaptive filter architectures for ECG signal denoising," *Biomedical Signal Processing and Control,* vol. 63, Jan 2021.
[30] T. S. Duan, W. K. Chen, M. L. Ruan, X. J. Zhang, S. F. Shen, and W. Y. Gu, "Unsupervised deep learning-based medical image registration: a survey," *Physics in Medicine and Biology,* vol. 70, Jan 19 2025.
[31] T. S. Huang, X. R. Huang, and H. B. Yin, "Deep learning methods for improving the accuracy and efficiency of pathological image analysis," *Science Progress,* vol. 108, Jan 2025.
[32] B. Pamukti, S. Afifah, S. K. Liaw, J. Y. Sung, and D. P. Chu, "Intelligent Pattern Recognition Using Distributed Fiber Optic Sensors for Smart Environment," *Sensors,* vol. 25, Jan 2025.
[33] Z. Liu, T. G. Liu, J. Zhao, J. Uduagbomen, Y. L. Wang, S. Popov, and T. H. Xu, "A Module to Enhance the Generalization Ability of End-to-End Deep Learning Systems in Optical Fiber Communications," *Journal of Lightwave Technology,* vol. 43, pp. 596-601, Jan 15 2025.
[34] J. Li, L. Huang, Y. K. Zhai, S. K. Ling, H. Ouyang, and L. Li, "MFF-IBL: A Lightweight Cascade Network Based on Multibranch Feature Fusion and Incremental Broad Learning for Lead-Independent ECG Classification," *Ieee Transactions on Instrumentation and Measurement,* vol. 74, 2025.
[35] P. Xia, Z. R. Bai, Y. C. Yao, L. R. Xu, H. Zhang, L. D. Du, X. X. Chen, Q. Ye, Y. S. Zhu, P. Wang, X. R. Li, G. Y. Wang, and Z. Fang, "Advanced Deep Neural Network with Unified Feature-Aware and Label Embedding for Multi-Label Arrhythmias Classification," *Tsinghua Science and Technology,* vol. 30, pp. 1251-1269, Jun 2025.
[36] J. L. Wang, R. Li, R. F. Li, B. Fu, C. X. Xiao, and D. Z. Chen, "Towards Interpretable Arrhythmia Classification With Human-Machine Collaborative

Knowledge Representation," *Ieee Transactions on Biomedical Engineering,* vol. 68, pp. 2098-2109, Jul 2021.

[37] O. Hussein, S. M. Jameel, J. M. Altmemi, M. A. Abbas, A. Ugurenver, Y. M. Alkubaisi, and A. H. Sabry, "Improving automated labeling with deep learning and signal segmentation for accurate ECG signal analysis," *Service Oriented Computing and Applications,* Nov 14 2024.

[38] Y. C. Wu, C. C. Lin, and C. Y. Yeh, "Severity Classification of Obstructive Sleep Apnea Using Electrocardiogram Signals," *Sensors and Materials,* vol. 36, pp. 4775-4780, 2024.

[39] Y. K. Guo, Q. F. Tang, S. Y. Li, and Z. C. Chen, "Reconstruction of Missing Electrocardiography Signals from Photoplethysmography Data Using Deep Neural Network," *Bioengineering-Basel,* vol. 11, Apr 2024.

[40] N. Murmu, R. Gupta, and K. Das Sharma, "Real-Time PPG-to-ECG Reconstruction Model With On-Device Recalibration Facility," *Ieee Transactions on Instrumentation and Measurement,* vol. 73, 2024.

[41] M. Das and B. C. Sahana, "A Deep-learning-based Auto Encoder-Decoder Model for Denoising Electrocardiogram Signals," *Iete Journal of Research,* Oct 8 2024.

[42] I. R. de Vries, J. O. E. H. van Laar, M. B. van der van der Jagt, and R. Vullings, "Unsupervised denoising of the non-invasive fetal electrocardiogram with sparse domain Kalman filtering and vectorcardiographic loop alignment," *Physiological Measurement,* vol. 45, Jul 1 2024.

[43] A. Pal, H. M. Rai, S. Agarwal, and N. Agarwal, "Advanced Noise-Resistant Electrocardiography Classification Using Hybrid Wavelet-Median Denoising and a Convolutional Neural Network," *Sensors,* vol. 24, Nov 2024.

[44] S. B. Gnanapirakasam and J. Manjula, "A Novel Approach for the Detection of Cardiovascular Abnormalities from Electrocardiogram and Phonocardiogram Signals Using Combined CNN-LSTM Techniques," *Traitement Du Signal,* vol. 41, pp. 3131-3142, Dec 2024.

[45] J. L. Wang, R. F. Li, R. Li, K. Q. Li, H. B. Zeng, G. Q. Xie, and L. Liu, "Adversarial de-noising of electrocardiogram," *Neurocomputing,* vol. 349, pp. 212-224, Jul 15 2019.

[46] P. Singh and G. Pradhan, "A New ECG Denoising Framework Using Generative Adversarial Network," *Ieee-Acm Transactions on Computational Biology and Bioinformatics,* vol. 18, pp. 759-764, Mar-Apr 2021.

[47] B. X. Xu, R. X. Liu, M. L. Shu, X. Y. Shang, and Y. L. Wang, "An ECG Denoising Method Based on the Generative Adversarial Residual Network," *Computational and Mathematical Methods in Medicine,* vol. 2021, Apr 20 2021.

[48] Y. Lan, H. Y. Xia, H. S. Li, S. X. Song, and L. Y. Wu, "Double Recurrent Dense Network for Single Image Deraining," *Ieee Access,* vol. 8, pp. 30615-30627, 2020.

[49] Y. V. Wang, S. H. Kim, G. Lyu, C. L. Lee, S. Ryu, G. Y. W. Lee, K. H. Min, and M. C. Kafatos, "Nowcasting Heavy Rainfall With Convolutional Long Short-Term Memory Networks: A Pixelwise Modeling Approach," *Ieee Journal of Selected Topics in Applied Earth Observations and Remote Sensing,* vol. 17, pp. 8424-8433, 2024.

[50] A. Sivaraj, P. Valarmathie, K. Dinakaran, and R. Rajakani, "Enhancing efficiency in agriculture: densely connected convolutional neural network for smart farming," *Signal Image and Video Processing,* vol. 18, pp. 6469-6480, Sep 2024.

[51] X. Shi, Z. Chen, H. Wang, D. Y. Yeung, W. K. Wong, and W. C. Woo, "Convolutional LSTM Network: A Machine Learning Approach for Precipitation Nowcasting," *MIT Press,* 2015.
[52] Z. Wang, A. C. Bovik, H. R. Sheikh, and E. P. Simoncelli, "Image quality assessment: From error visibility to structural similarity," *Ieee Transactions on Image Processing,* vol. 13, pp. 600-612, Apr 2004.
[53] Q. Huynh-Thu and M. Ghanbari, "Scope of validity of PSNR in image/video quality assessment," *Electronics Letters,* vol. 44, pp. 800-U35, Jun 19 2008.
[54] S. Qin, "Simple algorithm for L1-norm regularisation-based compressed sensing and image restoration," *Iet Image Processing,* vol. 14, pp. 3405-3413, Dec 2020.
[55] Y. M. Jin, X. H. Tong, L. Y. Li, S. L. Zhang, and S. J. Liu, "Total least L1-and L2-norm estimations of a coordinate transformation model with a structured parameter matrix," *Studia Geophysica Et Geodaetica,* vol. 59, pp. 345-365, Jul 2015.
[56] G. A. Moody and R. G. Mark, "The impact of the MIT-BIH arrhythmia database," *Ieee Engineering in Medicine and Biology Magazine,* vol. 20, pp. 45-50, May-Jun 2001.
[57] G. B. Moody, W. K. Muldrow, and R. G. Mark, "NOISE STRESS TEST FOR ARRHYTHMIA DETECTORS," *Computers in Cardiology,* vol. 11, pp. 381-384, 1984.
[58] A. Y. H. A. L. Maas, A. Y. Ng, "Rectifier nonlinearities improve neural network acoustic models," *2013 International Conference on Machine Learning,* vol. 30, p. 3, 2013.
[59] S. Ari, M. K. Das, and A. Chacko, "ECG signal enhancement using S-Transform," *Computers in Biology and Medicine,* vol. 43, pp. 649-660, Jul 1 2013.
[60] G. U. Reddy, M. Muralidhar, and S. Varadarajan, "ECG DeNoising using improved thresholding based on Wavelet transforms," *International journal of computer science and network security: IJCSNS,* p. 9, 2009.
[61] P. Xiong, H. R. Wang, M. Liu, S. P. Zhou, Z. G. Hou, and X. L. Liu, "ECG signal enhancement based on improved denoising auto-encoder," *Engineering Applications of Artificial Intelligence,* vol. 52, pp. 194-202, Jun 2016.